\documentclass[11pt]{article}
\usepackage{amssymb}
\usepackage{amsmath}
\usepackage{graphics,epsfig,color,soul}
\usepackage{bm}
\usepackage{color}
\usepackage{authblk}
\usepackage{wrapfig}
\usepackage{subcaption}
\usepackage{comment}
\usepackage{ragged2e}

\usepackage{tikz}
\usetikzlibrary{arrows.meta} 
\usetikzlibrary{decorations.pathreplacing,calligraphy}

\usepackage{cite}
\usepackage{hyperref}
\hypersetup{
  colorlinks   = true, 
  urlcolor     = blue, 
  linkcolor    = blue, 
  citecolor   = black 
}

\usepackage{framed} 
\newcommand{\tr}{\text{tr}}
\newcommand{\triplecontraction}{
  \mathrel{\vbox{\baselineskip.65ex\lineskiplimit0pt\hbox{.}\hbox{.}\hbox{.}}}
}

\begin{document}

\title{A computational study of nematic core structure and disclination
interactions in elastically anisotropic nematics}

\author{Lucas Myers$^{1}$, Carter Swift$^{1}$, Jonas
R\o{}nning$^{2}$, Luiza Angheluta$^{2}$, and Jorge Vi\~nals$^{2}$ \\
$^{1}$ School of Physics and Astronomy, University of Minnesota,
Minneapolis, MN 55455, USA. $^{2}$ Njord Center, Department of Physics,
Univesity of Oslo, P.O. Box 1048, 0316 Oslo, Norway}

\maketitle

\begin{abstract}

A singular potential method in the {\bf Q} tensor order parameter 
representation of a nematic liquid crystal is used to study the equilibrium 
configuration of a disclination dipole. Unlike the well studied isotropic limit
(the so called one constant approximation), we focus on the case of anisotropic
Frank elasticity (bend/splay elastic constant contrast). Prior research has
established that the singular potential method provides an accurate description
of the tensor order parameter profile in the vicinity of a disclination core of
a highly anisotropic lyotropic chromonic liquid crystal. This research is
extended here to two interacting disclinations forming a dipole configuration.
The director angle is shown to decay in the far field inversely with distance 
to the dipole as is the case in the isotropic limit, but with a diﬀerent 
angular dependence. Therefore elastic constant anisotropy modifies the elastic 
screening between disclinations, with implications for the study of
ensembles of defects as seen, for example, in active matter in the
extended system limit.
\end{abstract}

\section{Introduction}
\label{sec:introduction}

In nematic liquid crystals, the four distortion modes - splay, bend, twist, and
saddle splay - can each contribute differently to the elastic distortion
energy \cite{re:degennesde93,re:selinger16}, a phenomenon hereafter referred to
as \lq\lq anisotropic elasticity\rq\rq.  Even through the origin of this
anisotropic elasticity can be traced to the relative alignment of elongated
nematogens, and it is well documented, there still remain many open questions
related to the effects of anisotropic elasticity on the equilibrium and
nonequilibrium properties of defected nematics. A better understanding of the
role of anisotropy on the motion and interaction of disclinations is fundamental
to modeling biologically inspired and synthetic active matter systems.

In common thermotropic liquid crystals comprising small rod like molecules, the
contrast between splay, twist, and bend elastic constants is small, and the so
called one constant approximation has been successful in a wide variety of
applications. More recently, however, attention has shifted to systems comprised
of more complex nematogens which exhibit large elastic anisotropy. Chief among
them, we mention lyotropic chromonic liquid crystals
\cite{re:lydon10,re:park12,re:kim13,re:peng16,re:zhou17b} and nematic micellar
systems \cite{re:dietrich17,re:dietrich20}. Novel behavior has been uncovered
which is a direct result of elastic anisotropy, such spontaneously broken chiral
symmetry due to confinement
\cite{re:tortora11,re:nayani15,re:dietrich17,re:dietrich20,re:pellegrino21}, or
the existence and motion of topological solitons
\cite{re:li18,re:li20,re:tai20}. Complex anisotropic effects have also been
observed recently in studies of disclination line reconnection in three
dimensions \cite{re:zushi22}. 
In contrast with two dimensions, disclination lines in three dimensions only have a topological charge of 1/2, and can annihilate despite having the same charge sign. An apparent asymmetry in the motion of wedge disclination segments (of effective charge $\pm 1/2$) seems to be eliminated through twist in anisotropic media, thus restoring the implied topological symmetry.

The topology of defected configurations in two and three dimensional nematic
phases is well understood, including the case of biaxial ground states
\cite{re:alexander12}.  In two dimensions, the orientation $\theta(\mathbf r)$
of the nematic director $\hat{\mathbf{n}}$ is a harmonic function of $\mathbf r$
in the one constant (isotropic) approximation. Well known singular solutions are
associated with disclination point sources \cite{re:chaikin95,re:selinger16}.
Configurations comprising many disclinations can be described by linear
superposition, and results have been given for a number of cases of interest,
including, for example, binding-unbinding transitions in active matter
\cite{re:shankar18}, or defect interactions in complex twisted configurations
obtained by conformal mapping techniques \cite{re:tang17}. In contrast, little
is known about nematic director $\hat{\mathbf{n}}$ or tensor order parameter
$\mathbf{Q}$ configurations corresponding to defected configurations in
elastically anisotropic media, both in two and three dimensions. A key result in
two dimensions was obtained by Dzyaloshinskii
\cite{re:dzyaloshinskii70,re:hudson89}. When the splay $K_{1}$ and bend $K_{3}$
elastic constants are different, he found an analytic -albeit only implicit-
solution for the equilibrium nematic orientation $\theta$ corresponding to an
isolated disclination. The solution is independent of distance from the core,
but depends on the azimuthal angle around the disclination. More generally, the
Euler-Lagrange equations that follow from the Frank free energy are nonlinear
and challenging to solve analytically. 

While it is possible to study both equilibrium and transient configurations of
nematics containing disclinations in the director representation, with the Frank
free energy governing elastic distortion, and Leslie-Ericksen hydrodynamics
fluid flow, it is often the case that a $Q$ tensor order parameter
representation and the Landau-de Gennes energy are used instead. Virtually all
studies of nematic active and biological matter use this representation as it
eliminates the need for defect core regularization (specially in three
dimensions), and hence it permits a more convenient computational treatment of
disclinations and their motion. Unfortunately, this choice has the effect of
restricting these studies to the one constant approximation. Elasticity in the
tensor order parameter representation is incorporated in a phenomenological
series expansion in powers of order parameter gradients, Eq. \eqref{eq:elastic}
below. For small distortions, Frank elastic constants can be related to the
coefficients of the expansion as shown in Eq.
\eqref{eq:frank-ldg-elastic-comparison}. In order to capture splay-bend
anisotropy, the lowest order coefficient that needs to be nonzero, $L_{3}$,
corresponds to a cubic term in the order parameter. At this order, the Landau-de
Gennes energy is known to become unbounded for any choice of parameters
\cite{re:ball10,re:bauman16}. Unboundedness can be traced back to the lack of
any constraint in the Landau-de Gennes free energy on the physical admissibility
of the eigenvalues of $\mathbf{Q}$. In principle, the requirement of a stable
free energy could be accomplished in this case by consideration in the expansion
defining ${\cal F}_{el}$ of terms at least of fourth order in $\mathbf{Q}$
\cite{re:koizumi23}. However, since there are 22 possible invariants up to
fourth order allowed by symmetry \cite{re:longa87}, the Landau-de Gennes theory
becomes intractable for an elastically anisotropic nematic. Building into the
theory the constraint that the eigenvalues of $\mathbf{Q}$ must remain within
the physically admissible range can be accomplished by an appropriately defined
singular potential
\cite{re:katriel86,re:ball10,re:schimming20a,re:schimming20b,re:schimming21}.
The drawback of this theory is that the determination of the energy needs to be
done entirely numerically at a significant computational cost relative to simple
evaluations of the Landau-de Gennes energy.

Two complementary issues are investigated below in relation to elastically
anistropic nematic phases, both in the tensor order parameter representation.
First, we build on the singular potential method analysis of Ref.
\cite{re:schimming20b} to quantitatively describe both bialixiality and
anisotropy of disclination cores. We use the method to compute the optical
retardance, $\Gamma = S - P$, near a disclination core, where $S$ and $P$ are
the uniaxial and biaxial order parameters respectively. Exactly at the
disclination core, $S = P$, in agreement with experiments \cite{re:zhou17} and
earlier calculations \cite{re:schimming20b}. We then show that as the core is
approached $\Gamma_{0} \sim r$, with $\Gamma_{0}$ being the isotropic component
of the angular Fourier transform of $\Gamma$, and $r$ the radial distance from
the core. We also show that anisotropic azymuthal components $\Gamma_{1}$ for a
+1/2 disclination and $\Gamma_{3}$ for a -1/2 disclination are nonzero in the
biaxial region. However, they vanish as $r^{2}$ as the core is approached.
Hence, the uniaxial and anisotropic far field leads an anisotropic and biaxial
region as the core is approached. At even smaller distances, the configuration
becomes both uniaxial and isotropic, as judged from the azymuthal Fourier
transform of $\Gamma$.

Second, we focus on the interaction of a pair of disclinations of opposite sign
(a disclination dipole), and examine the nature of their screening at distances
much larger than their separation. For isotropic elasticity, the orientation
angle far from the disclination pair behaves as $\theta = q_{1} + q_{2} -
d(q_{1}-q_{2}) \sin \varphi/(2r)$ where $q_{1,2}= \pm 1/2$ are the charges of
the disclinations separated by distance $d$, $r$ is the radial distance from the
pair, and $\varphi$ is the azymuthal angle measured relative to the separation
distance vector. For two disclinations of opposite charge, the distortion is
screened and decays algebraically as $1/r$, modulated by $\sin \varphi$ in
angular dependence. In the anisotropic case, the far field dependence contains
an additional term of the form $\pm d \sin (3 \varphi)/r$ which has the same
decay with distance, but a different angular dependence. As a consequence,
disclination interactions in elastically anisotropic nematics are qualitatively
different than their isotropic counterparts, and the implications of these
findings on current phenomenology involving multiple defect interactions and
motion need to be reexamined.

\section{Nematic director and $\bf{Q}$ tensor representations} \label{director-model}

The description of the nematic phase of a liquid crystal introduces the local direction of orientational order, the director field, a unit vector $\mathbf{n}(\mathbf{x})$. This field corresponds to the average orientation direction of the constituent molecules, with configurations being invariant under the transformation $\mathbf{n} \to -\mathbf{n}$. A free energy penalizing distortions away from a uniform ground state is introduced containing all scalar combinations of gradients of $\mathbf{n}$ to second order that respect $\mathbf{n} \to -\mathbf{n}$. The resulting Frank free energy reads \cite{re:selinger16},
\begin{equation}
    F_e(\mathbf{n}, \nabla \mathbf{n}) =
    \int_\Omega \biggl[
    \frac12 K_1 (\nabla \cdot \mathbf{n})^2
    + \frac12 K_2 \left[ \mathbf{n} \cdot (\nabla \times \mathbf{n}) \right]^2 
    + \frac12 K_3 \left| \mathbf{n} \times (\nabla \times \mathbf{n}) \right|^2
    + \frac12 K_{24} \nabla \cdot \left[(\mathbf{n} \cdot \nabla) \mathbf{n} - \mathbf{n} (\nabla \cdot \mathbf{n})\right]
    \bigg]
    dV
\end{equation}
with the $K_1, K_2, K_3, K_{24}$ terms representing the splay, twist, bend, and
saddle splay distortion modes respectively. In two dimensions, the twist and
saddle-splay terms are manifestly zero. Introduce an anisotropy parameter
$\epsilon = (K_3 - K_1) / (K_3 + K_1)$, dimensionless length units $\overline{x}
= x / \xi$, and dimensionless free energy $\overline{F_n} = 2 F_n / (K_1 +
K_3)$. Dropping the overlines for simplicity one finds, \begin{equation}
\label{eq:2D-frank-energy}
    F_n(\mathbf{n}, \nabla \mathbf{n})
    =
    \int_\Omega
    \left[
    (1 - \epsilon) (\nabla \cdot \mathbf{n})^2
    + (1 + \epsilon) \left| \mathbf{n} \times (\nabla \times \mathbf{n}) \right|^2
    \right]
    dV
\end{equation}

The minimizer of Eq. \eqref{eq:2D-frank-energy} for a single point disclination in an infinite medium and for arbitrary $\epsilon$ has been given by Dzyaloshinskii, though only implicitly as an integral equation \cite{re:dzyaloshinskii70,re:hudson89}. The nematic director $\mathbf{n} = (\cos \theta, \sin \theta)$ is determined by the orientation field $\theta$, which is independent of the distance $r$ from the point defect, and depends only the azimuth $\varphi$, i.e. $\theta(\varphi)$. The Euler-Lagrange equation describing the minimizer of the Frank free energy in Eq. \eqref{eq:2D-frank-energy} is then \begin{equation} \label{eq:frank-polar-euler-lagrange}
    \frac{d^{2}\theta}{d \varphi^{2}}
    =
    \epsilon \biggl[ \frac{d^{2}\theta}{d \varphi^{2}} \cos 2 (\theta - \varphi) 
    + \left( 2 \frac{d \theta}{d \varphi} - \left( \frac{d \theta}{d \varphi}\right)^{2} \right) \sin 2 (\theta - \varphi)  .
    \biggr]
\end{equation}
In the isotropic limit $\epsilon = 0$, the director orientation is multivalued and given by $\theta_\text{iso} (\varphi) = q \varphi$, where $q= \pm 1/2$ is the disclination charge~\cite{re:zhou17}. A perturbative solution in $\epsilon$ can be found by expanding,
\begin{equation}
\label{eq:theta-perturbative-expansion}
    \theta (\varphi)
    =
    \theta_\text{iso}(\varphi)
    + \epsilon \theta_c (\varphi)
    + \mathcal{O}(\epsilon^2),
\end{equation}
where the first order correction is nonlinear in $\varphi$ \cite{re:zhou17}
\begin{equation} \label{eq:isolated-defect-perturbative-solution}
    \theta_c
    =
    \frac{q (2-q)}{4 (1-q)^{2}}\sin (2 (1-q) \varphi).
\end{equation}

In order to capture both the magnitude and orientation of nematic uniaxial/biaxial order, one introduces a tensor order parameter $\mathbf{Q}$  which is a coarse-grained, statistical measure of the nematic alignment defined by the second moment
\begin{equation} \label{eq:Q-tensor-definition}
    \mathbf{Q}
    =
    \int_{S^2}
    \left( \mathbf{p} \otimes \mathbf{p} - \tfrac13 I \right) \, \rho(\mathbf{p}) \, dS(\mathbf{p}).
\end{equation}
Here $\rho(\mathbf{p})$ is the probability density function of molecular orientation $\mathbf{p}$ defined on $S^2$, the unit sphere. Because of nematic symmetry, one has $\rho(\mathbf{p}) = \rho(-\mathbf{p})$. Since $\mathbf{Q}$ is traceless and symmetric by definition, its three eigenvectors $\mathbf{n}, \mathbf{m}, \mathbf{l}$ form an orthonormal basis, so that $\mathbf{Q}$ may be written as:
\begin{equation}
  \mathbf{Q}
  = S (\mathbf{n} \otimes \mathbf{n} - \tfrac13 I)
  + P (\mathbf{m} \otimes \mathbf{m} - \mathbf{l} \otimes \mathbf{l})
\end{equation}
The scalar order parameter $S$ describes the degree to which molecules are aligned along the director $\mathbf{n}$, while $P$ describes biaxiality, or the difference in alignment along the two remaining axes.

A Landau-de Gennes free energy expansion is usually introduced in terms of scalar contractions of $Q$ (the \lq\lq bulk" terms), supplemented by terms in gradients of $Q$ (the \lq\lq elastic" terms). For small distortion and fixed $S$, the elastic terms in the Landau-de Gennes free energy may be mapped onto the Frank elastic free energy exactly. However, in order to include bend-splay anisotropy, one must expand the elastic energy up to third order in gradients of $\mathbf{Q}$. It is well known that such a free energy is unbounded below \cite{re:ball10,re:bauman16}, although fourth order terms can be introduced to render the free energy finite \cite{re:koizumi23}. However, there are 22 possible terms allowed by symmetry up to fourth order, resulting in a proliferation of phenomenological coefficients that makes the theory unwieldy.

In order to overcome this difficulty, the Ball-Majumdar singular bulk potential method is introduced \cite{re:ball10,re:schimming21}. One defines a bulk free energy $F_{b}[\mathbf{Q}] = E[\mathbf{Q}] - T \Delta {\cal S}[\mathbf{Q}]$ where $E$ is the bulk energy, $T$ is the temperature, and $\Delta {\cal S}$ is the entropy relative to the isotropic phase.  The energy is chosen to be of the Maier-Saupe form $E[\mathbf{Q}] = -\kappa \int_\Omega \tr\left[\mathbf{Q}(\mathbf{r})\right] dV$ where $\kappa$ is a positive constant that characterizes alignment strength.  The entropy may be written in terms of the molecular probability distribution function,
\begin{equation} \label{eq:entropy-definition}
    \Delta {\cal S}
    =
    -n k_B \int_\Omega \int_{S^2} \rho(\mathbf{p}, \mathbf{r})
    \ln \left[4\pi \rho(\mathbf{p}, \mathbf{r}) \right]
    dS(\mathbf{p})
    dV
\end{equation}
where $n$ is the number density of nematogens, $k_B$ is Boltzmann's constant, and the probability density function of molecular orientation $\rho$ is allowed to be a function of position for an inhomogeneous configuration. In order to find an explicit expression of $\Delta {\cal S}$ in terms of $\mathbf{Q}$, \; $\rho$ is determined so that it maximizes $\Delta {\cal S}$ subject to the constraint \eqref{eq:Q-tensor-definition}.
The solution is,
\begin{equation} \label{eq:PDF-solution}
    \rho(\mathbf{p})
    =
    \frac{\exp \left( \mathbf{p}^T \bm{\Lambda} \mathbf{p} \right)}
    {Z[\bm{\Lambda}]}
\end{equation}
with partition function $Z$ given by:
\begin{equation}
    Z[\bm{\Lambda}] 
    =
    \int_{S^2} \exp\left(\mathbf{p}^T \bm{\Lambda} \mathbf{p}\right) dS(\mathbf{p}),
  \end{equation}
  where $\bm{\Lambda}$ is a tensor of Lagrange multipliers arising from the constraint \eqref{eq:Q-tensor-definition}. By substituting Eq. \eqref{eq:PDF-solution} into Eq. \eqref{eq:Q-tensor-definition} we may relate the multipliers $\bm{\Lambda}$ to $\mathbf{Q}$ as a mean field consistency condition,
\begin{equation} \label{eq:singular-potential-solution}
    \mathbf{Q}
    =
    \frac{\partial \ln Z}{\partial \bm{\Lambda}}
    - \frac13 \mathbf{I}.
\end{equation}
Substituting Eq. \eqref{eq:PDF-solution} into Eq. \eqref{eq:entropy-definition} and using Eq. \eqref{eq:singular-potential-solution} to simplify, the entropy may be written in terms of $\mathbf{Q}$ as,
\begin{equation}
    \Delta {\cal S}
    =
    -n k_B \int_\Omega
    \left[ \ln 4 \pi - \ln Z[\mathbf{Q}] + \bm{\Lambda}[\mathbf{Q}] : \left(\mathbf{Q} + \tfrac13 \mathbf{I} \right) \right]
    dV
\end{equation}
where $:$ is a double index contraction.  

For the elastic free energy, we include one term of third order in $\mathbf{Q}$ to allow for bend-splay anisotropy, 
\begin{equation} \label{eq:LdG-elastic-free-energy}
    F_\text{el}
    =
    \int_\Omega
    \biggl[
        L_1 \left| \nabla \mathbf{Q} \right|^2
        + L_2 \left| \nabla \cdot \mathbf{Q} \right|^2
        + L_3 \left( \nabla \mathbf{Q} \right) \triplecontraction \left[ \left( \mathbf{Q} \cdot \nabla \right) \mathbf{Q} \right]
    \biggr] dV
\end{equation}
where $\triplecontraction$ is a triple index contraction from inner indices to outer indices. Written in index notation this equation reads,
\begin{equation}
    F_\text{el} [\mathbf{Q}, \nabla \mathbf{Q}]
    =
    \int_\Omega
    \biggl[
        L_1 \left( \partial_k Q_{ij} \right)^2
        + L_2 \left( \partial_j Q_{ij} \right)^2
        + L_3 Q_{lk} \left( \partial_{l} Q_{ij} \right) \left( \partial_k Q_{ij} \right)
    \biggr] dV
\label{eq:elastic}
\end{equation}
We recall that the mapping to the Frank free energy coefficients in the case of a uniaxial and constant $S$ nematic phase is given by \cite{re:mottram14}:
\begin{equation} \label{eq:frank-ldg-elastic-comparison}
\begin{split}
    K_1 &= 4 L_1 S^2 + 2 L_2 S^2 - \tfrac43 L_3 S^3 \\
    K_2 &= 4 L_1 S^2 - \tfrac43 L_3 S^3 \\
    K_3 &= 4 L_1 S^2 + 2 L_2 S^2 + \tfrac83 L_3 S^3 \\
    K_{24} &= 4 L_1 S^2 - \tfrac43 L_3 S^3
\end{split}
\end{equation}
The total free energy in the singular potential method is the sum $F = F_{b} + F_\text{el}$.

Rotational relaxation dynamics of the nematogens is considered through
\begin{equation} \label{eq:overdamped-equation-of-motion}
    \frac{\partial \mathbf{Q}}{\partial t} = - \gamma \frac{\delta F}{\delta \mathbf{Q}}.
\end{equation}
We introduce the dimensionless quantities:
\begin{equation}
    \overline{x} = x / \xi, \:\:\:\:
    \overline{t} = t / \tau, \:\:\:\:
    \overline{\kappa} = \frac{2 \kappa}{n k_B T}, \:\:\:\:
    \overline{L_2} = \frac{L_2}{L_1}, \:\:\:\:
    \overline{L_3} = \frac{L_3}{L_1}
\end{equation}
where the rescaling units of length and time are given by 
\begin{equation} \label{eq:nondimensional-length-time}
    \xi = \sqrt{\frac{2 L_1}{n k_B T}}, \:\:\:\:
    \tau = \frac{1}{\gamma \, n k_B T}
\end{equation}
Dropping the overlines for simplicity, the dimensionless equation of motion for $\mathbf{Q}$ is,
\begin{eqnarray} \label{eq:Q-tensor-equation-of-motion}
\frac{\partial \mathbf{Q}}{\partial t} & = & \kappa \mathbf{Q} - \bm{\Lambda}
      + \nabla^2 \mathbf{Q}  \nonumber \\
      & + & L_2 \left( \nabla \left( \nabla \cdot \mathbf{Q} \right)
        + \left[ \nabla \left( \nabla \cdot \mathbf{Q} \right) \right]^T
        - \tfrac23 \left( \nabla \cdot \left( \nabla \cdot \mathbf{Q} \right) 
\right) \mathbf{I}
      \right) \\
      & + & L_3 \left(
        2 \nabla \cdot \left( \mathbf{Q} \cdot \nabla \mathbf{Q} \right)
        - \left( \nabla \mathbf{Q} \right) : \left( \nabla \mathbf{Q} \right)^T
        + \tfrac13 \left| \nabla \mathbf{Q} \right|^2 \mathbf{I}
      \right) \nonumber
\end{eqnarray}
with the transpose of a rank 3 tensor being defined as $(\nabla
\mathbf{Q})^{T}_{klj} = \partial_{j}Q_{kl}$.
Hereafter, all distances and times will be in units of $\xi$ and $\tau$ respectively.

Equilibrium configurations correspond to $\partial_t\mathbf{Q} = 0$. The resulting nonlinear elliptic pde may be solved numerically by using the Newton-Rhapson relaxation method. This method is used below for configurations with an isolated disclination. For the case of a disclination pair, however, the Newton-Rhapson method is not computationally efficient due it to its slow convergence for large systems. Instead we discretize Eq. \eqref{eq:Q-tensor-equation-of-motion} in time by using a Crank-Nicolson method. We then use the same Newton-Rhapson method to solve for each subsequent time step, and iterate in time until $\partial_t \mathbf{Q}$ is sufficiently small. The Appendices provide additional details.

Boundary conditions in a finite domain need to be discussed separately. Given the variational derivative of the energy $\frac{\delta F}{\delta \mathbf{Q}} = \frac{\partial f}{\partial \mathbf{Q}} - \nabla \cdot \frac{\partial f}{\partial \left(\nabla \mathbf{Q} \right)}$, we impose Neumann boundary conditions by requiring that the normal component at the outer boundary $\mathbf{N} \cdot \partial f / \partial \left( \nabla Q \right) = 0$, where $\mathbf{N}$ is the outward pointing normal. This reduces to the familiar Neumann boundary condition on $\mathbf{Q}$ in the isotropic limit, but more generally, it is the natural boundary condition to use for a fully anisotropic system. 

\section{A single disclination in the {\bf Q} tensor representation} 
\label{sec:isolated-disclination-core}

We address first the asymptotic dependence of the $\mathbf{Q}$ tensor near the core of an isolated disclination in an elastically anisotropic medium. This short length scale structure has been experimentally characterized in lyotropic chromonics \cite{re:zhou17}, and shown theoretically to determine the kinematics of disclination motion \cite{re:schimming22,re:schimming23}. 
\begin{figure*}[t]
\captionsetup[subfigure]{justification=Centering}

  \hfill
  \begin{subfigure}{0.35\textwidth}
    \includegraphics[width=\textwidth]{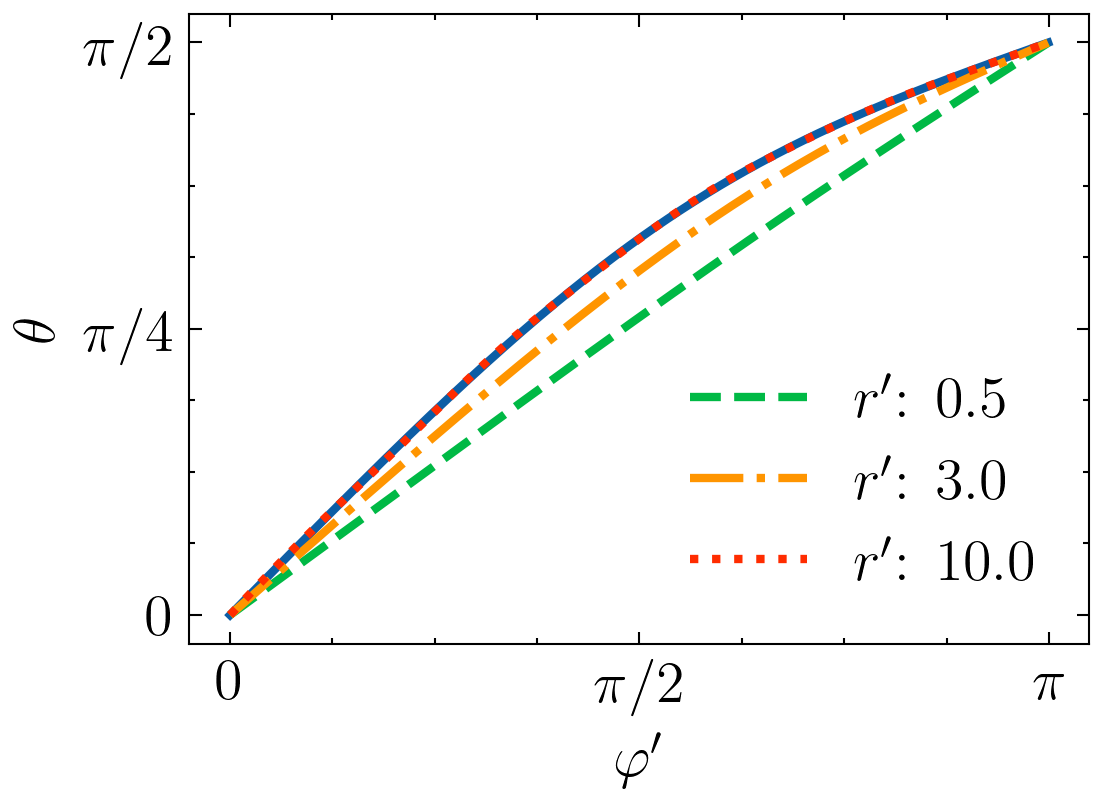}
    \caption{}
    \label{fig:plus-half-disclination-director-angle}
  \end{subfigure}
  \hfill
  \begin{subfigure}{0.47\textwidth}
    \includegraphics[width=\textwidth]{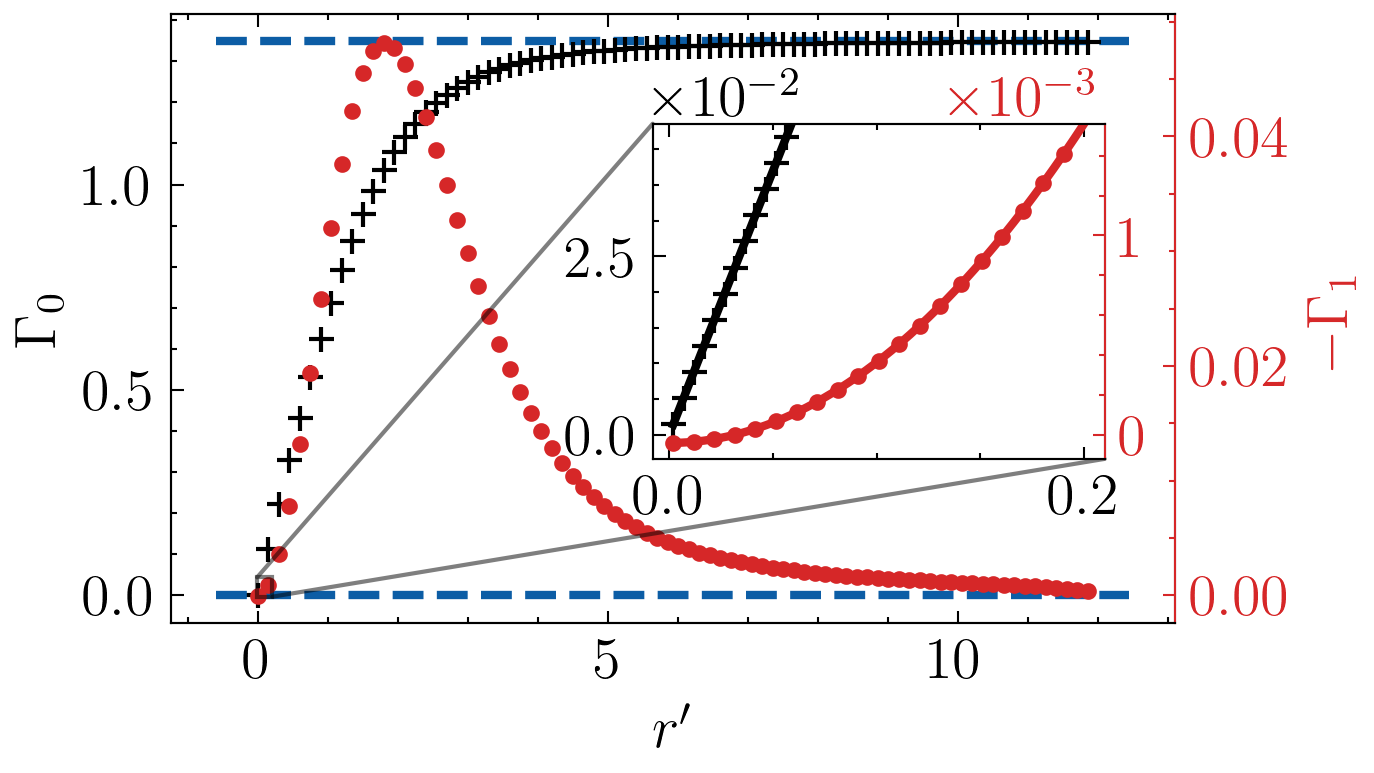}
    \caption{}
    \label{fig:plus-half-isolated-disclination-S_P}
  \end{subfigure}
  \hfill

  \hfill
  \begin{subfigure}{0.35\textwidth}
    \includegraphics[width=\textwidth]{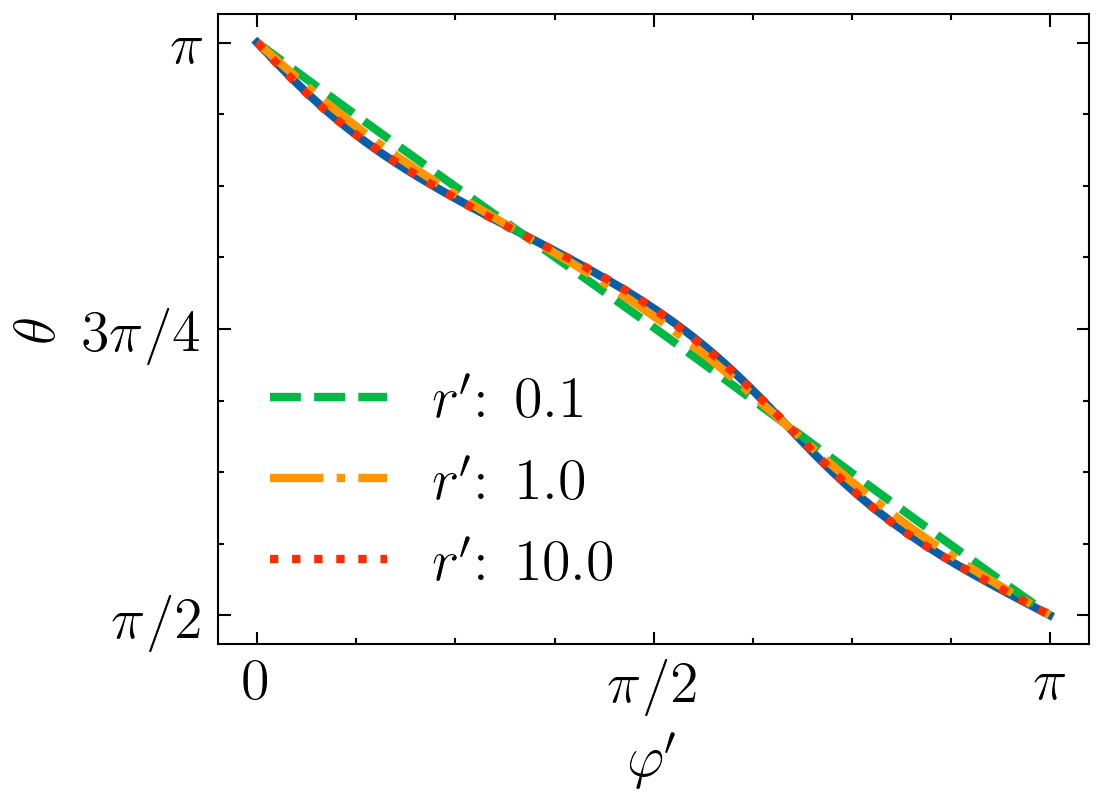}
    \caption{}
    \label{fig:minus-half-disclination-director-angle}
  \end{subfigure}
  \hfill
  \begin{subfigure}{0.47\textwidth}
    \includegraphics[width=\textwidth]{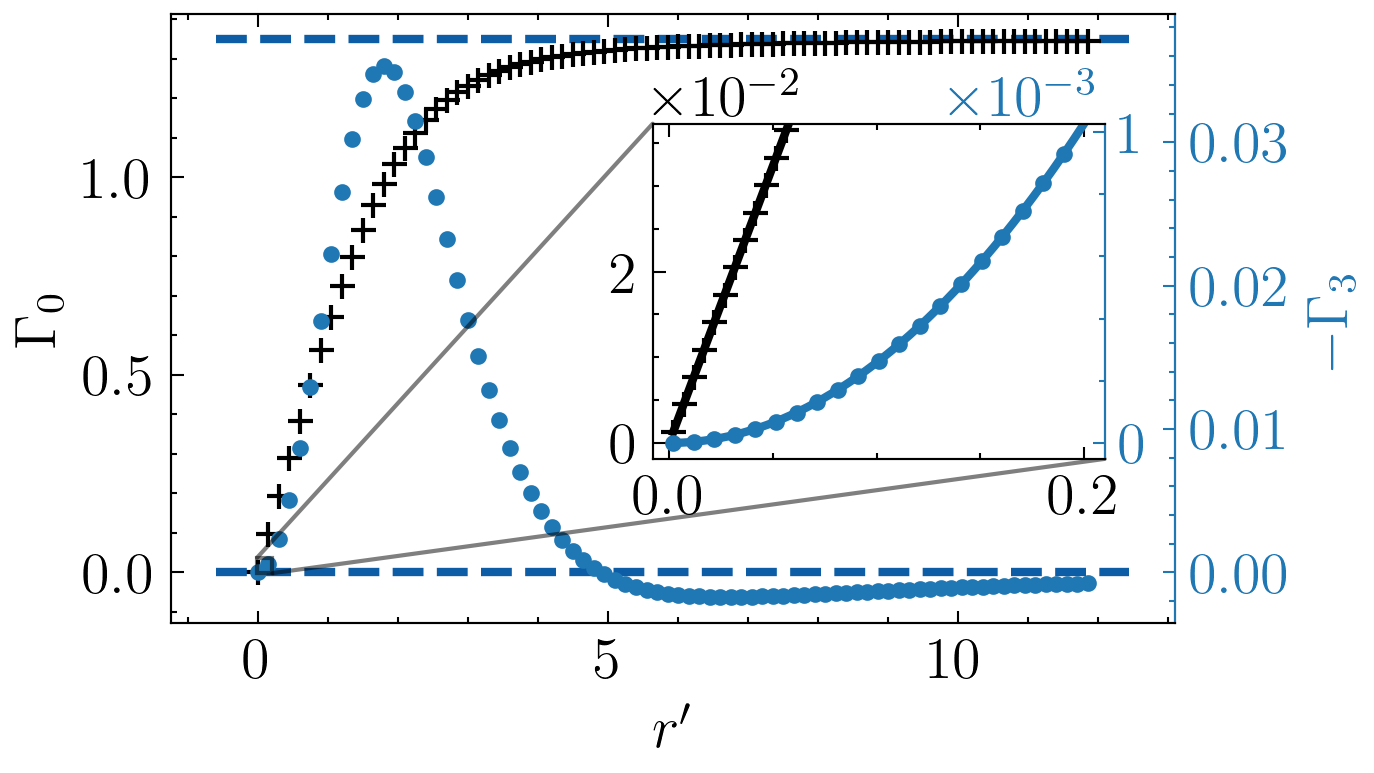}
    \caption{}
    \label{fig:minus-half-isolated-disclination-S_P}
  \end{subfigure}
  \hfill
    \caption{(a), (c) Director angle $\theta$ as a function of the azymuth
$\varphi'$ at various distances from the core for $+1/2$ and $-1/2$
disclinations respectively, computed from the equilibrium $\mathbf{Q}$ tensor.
The solid line is $\theta_\text{DZ}({\rm atan2}(r'\sin\varphi' + y_\text{disc}, r'\cos\varphi' + x_\text{disc}))$ with $\theta_\text{DZ}(\varphi)$ the solution to Eq. \eqref{eq:frank-polar-euler-lagrange} and $(x_\text{disc}, y_\text{disc})$ the disclination centers. (b), (d) Angular Fourier decomposition of $\Gamma$ as a function of distance from the defect core for $+1/2$ and $-1/2$ disclinations respectively. The insets shows the asymptotic behavior as the disclination core is approached. Pluses ($\Gamma_0$) and dots ($\Gamma_1$, $\Gamma_3$) are points obtained from the numerical solutions, dashed horizontal lines represent the long distance equilibrium values of $S = S_0$ (and $P=0$), and solid lines are fits of the form $A (r')^n + B$. Fit coefficients for the $+1/2$ disclination are $A = 0.733$, $n = 0.996$, $B = -8.69\times 10^{-5}$ and $A = 0.0392$, $n = 1.986$, $B = -4.23 \times 10^{-5}$ for $\Gamma_0$ and $\Gamma_1$ respectively. Fit values for the $-1/2$ disclination are $A = 0.644$, $n = 0.998$, $B = -3.95\times 10^{-5}$ and $A = 0.0253$, $n = 1.990$, $B = -3.26 \times 10^{-7}$ for $\Gamma_0$ and $\Gamma_3$ respectively. We note that the data points shown in the figure are only a small subset as our numerical solution has a resolution of $r \approx 0.002$.}
  \label{fig:isolated-disclination-core-structure}
\end{figure*}
A numerical calculation is carried out in a two dimensional circular domain of radius $R = 20 / \sqrt{2}$ to compute the equilibrium nematic configuration around an isolated $\pm 1/2$ disclination for dimensionless values of the parameters  $\kappa = 8.0$, $L_2 = 4.58$, $L_3 = 4.5$. This value of $\kappa$ corresponds to an equilibrium value of $S$ to be $S_0 = 0.6751$. 
The thin film approximation is used for $\mathbf{Q}$ so that the tensor is described by three independence components, not just two as for a strictly two dimensional case. The domain is discretized with quadrilaterals, initially with $12$ cells. 
It is then globally refined $5$ times, and further refined at distances $R = 8, 4, 2, 1, \frac{1}{2}, \frac{1}{4}, \frac{1}{8}, \frac{1}{16}$ from the initial disclination center. Every refinement operation divides each quadilateral cell into four children cells. 
The stationary disclination cores are located at $(0, 0)$ and $(0.868, 0)$ for the $-1/2$ and $+1/2$ disclinations respectively. 
We label these disclination core locations as $(x_\text{disc}, y_\text{disc})$ and define polar coordinates $(r', \varphi')$ centered at these coordinates.
In this calculation, we have imposed Dirichlet boundary conditions on $\mathbf{Q}$ on the outer boundary as a uniaxial, $\mathbf{Q}$ tensor configuration, with $S = S_0$, and director angle equal to the numerical solution to Eq. \eqref{eq:frank-polar-euler-lagrange}, with $\epsilon$ obtained from $S_0$, $L_{1}$, and $L_3$, via Eq. \eqref{eq:frank-ldg-elastic-comparison}, and polar angle $\varphi$ centered at the domain origin.

We find the director $\mathbf{n}$ and scalar order parameters $S$ and $P$ by calculating the eigenvalues and corresponding eigenvectors of the $\mathbf{Q}$ tensor at each point. This is done with the \verb|eigh| method from the Numpy numerical package, which calculates the eigensystem of a symmetric matrix \cite{re:numpy}. Let $\lambda_1$ and $\lambda_2$ be the largest and second largest eigenvalues of $\mathbf{Q}$ respectively, then $S = \frac32 \lambda_1$ and $P = \frac12 \lambda_1 + \lambda_2$. The nematic director is the eigenvector corresponding to the largest eigenvalue.

The disclination core in the $\mathbf{Q}$ tensor representation is located at $S=P$ (the optical retardance measured experimentally is $\Gamma \propto S-P$). Away from the core, any azimuthal angular dependence of $\Gamma$ is a measure of elastic anisotropy.  We therefore examine the anisotropy through its Fourier series decomposition 
\begin{equation} \label{eq:S-P-fourier-expansion}
    \Gamma(r', \varphi') = \sum_n \Gamma_n (r') \sin (n \varphi')
\end{equation}
Numerically, the Fourier coefficients are calculated with the \verb|rfft| real Fourier transform method from the Numpy numerical package. The sine coefficients in Eq. \eqref{eq:S-P-fourier-expansion} are $2 / N$ times the real part of the discrete transform modes, where $N$ is the number of discretized points at each $r'$ \cite{re:numpy}.

Figures \ref{fig:plus-half-disclination-director-angle} and \ref{fig:minus-half-disclination-director-angle} show the director angle $\theta$ vs. the azimuth $\varphi'$ plotted at several fixed distances from the disclination centers. We note that the core of the $+1/2$ disclination is located at $(0.868, 0)$ in the equilibrium configuration, slightly offset from the domain center. 
At large distances, the director profile approaches the Dzyaloshinskii perturbative solution of Eq. \eqref{eq:frank-polar-euler-lagrange} calculated at the domain center, but plotted as a function of $\varphi'$ at $r' = 10$.
Explicitly, if $\theta_\text{DZ}(\varphi)$ is the solution to Eq.
\eqref{eq:frank-polar-euler-lagrange}, the solid line in Fig.
\ref{fig:plus-half-disclination-director-angle} is given by
$\theta_\text{DZ}({\rm atan2}(r'\sin\varphi' + y_\text{disc}, r'\cos\varphi' + x_\text{disc}))$ for $r' = 10$. 
However, as $r' \to 0$, the director tends towards the isotropic solution $\theta = \tfrac12 \varphi'$. Figures \ref{fig:plus-half-isolated-disclination-S_P} and \ref{fig:minus-half-isolated-disclination-S_P} show the two dominant angular Fourier modes $\Gamma_n(r')$ as a function of radial distance to the defect core. The figures also show the fit to an asymptotic power law with distance. The zeroth Fourier modes goes to zero linearly, while the higher Fourier modes appear to decrease quadratically as the core is approached.

The dependence of the eigenvalues near disclination cores has been previously studied numerically for isotropic elasticity using a Landau-de Gennes bulk free energy \cite{re:schopohl87}. The director configuration is azimuthally symmetric, and the two largest eigenvalues change linearly as the core is approached \cite{re:schopohl87}. Recent work has used these facts to approximate the core eigenvalue structure as a piecewise linear function of $r'$, allowing an analytic investigation of disclination orientation \cite{re:long21}. Further, this ``linear core approximation'' has been used along with the disclination kinematic law to predict disclination motion in \textit{anisotropic} media \cite{re:schimming22, re:schimming23}. Our results show that the amplitudes of anisotropic Fourier modes vanish faster than the isotropic zeroth order mode; hence the director angle approaches that of a disclination in an isotropic medium as the core is approached. Furthermore, the dominant dependence of the eigenvectors is also linear as the core is approached, in agreement with the isotropic results. Both observations suggest that the linear core approximation is a reasonable approximation even in anisotropic media. The complicating factor that remains, and to which we turn next, is that in two or multi defect configurations, the tensor field is not a superposition of configurations corresponding to isolated single defects. Therefore it remains to be seen whether interaction leads to a more complicated core structure in multi disclination systems.

\section{A disclination dipole}
\label{sec:perturbative_solution}

\begin{figure} 
    \centering
    \begin{tikzpicture}[auto, >={Stealth[width=2.0]}]
        \coordinate (q1) at (-2, 0);
        \coordinate (q2) at (2, 0);
        \coordinate (x) at (2.5, 2.0);
        \coordinate (o) at (0, 0);
        \coordinate (q1x) at (-0.5, 0);
        \coordinate (q2x) at (3.2, 0);
        \coordinate (ox) at (1.2, 0);

        \fill (q1) circle [radius=0.04];
        \node[below, left] at (q1) {$q_1$};
        \fill (q2) circle [radius=0.04];
        \node[below, left] at (q2) {$q_2$};
        \fill (o) circle [radius=0.04];

        \draw[->] (q1) -- node {$r_1$} (x);
        \draw[->] (q2) -- node [swap, right] {$r_2$} (x);
        \draw[->] (o) -- node [swap] {$r$} (x);
        \draw[dashed] (q1) -- (q1x);
        \draw[dashed] (q2) -- (q2x);
        \draw[dashed] (o) -- (ox);

        \begin{scope}
            \clip (q1x) -- (q1) -- (x) -- cycle;
            \draw (q1) circle [radius=0.5];
            \draw ([shift={(q1)}] 13:1.0) node {$\varphi_1$};
        \end{scope}

        \begin{scope}
            \clip (q2x) -- (q2) -- (x) -- cycle;
            \draw (q2) circle [radius=0.5];
            \draw ([shift={(q2)}] 33:0.8) node {$\varphi_2$};
        \end{scope}

        \begin{scope}
            \clip (ox) -- (o) -- (x) -- cycle;
            \draw (o) circle [radius=0.5];
            \draw ([shift={(o)}] 20:0.8) node {$\varphi$};
        \end{scope}

        \draw [decoration={brace,mirror, raise=5pt}, decorate] (q1) -- node[below=7pt] {$d$} (q2) ;
    \end{tikzpicture}
    \caption{Diagram showing a disclination pair in polar coordinates. Here $(r_i, \varphi_i)$ are polar coordinates centered on the disclination with charge $q_i$, and $(r, \varphi)$ are polar coordinates centered on the midpoint between the to disclinations.}
    \label{fig:perturbative-disclination-pair-polar-coordinates}
\end{figure}
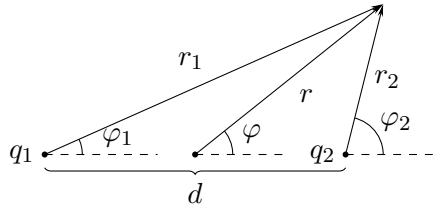

The Euler-Lagrange equations corresponding to the Frank energy \eqref{eq:2D-frank-energy} in Cartesian coordinates read,
\begin{equation} \label{eq:frank-cartesian-euler-lagrange}
    \nabla^{2} \theta 
    = \epsilon \left[ \sin (2 \theta) (\theta_{x}^{2} - \theta_{y}^{2}-2 \theta_{xy}) 
        + \cos (2 \theta) (\theta_{yy}-\theta_{xx}-2\theta_{x}\theta_{y})\right] 
\end{equation}
Consider now a pair of disclinations a distance $d$ from each other, which are mutually aligned or anti aligned. We seek a perturbative solution for the director field to first order in $\epsilon$ \cite{re:swift22t,re:swift22}. The solution in the isotropic limit of $\epsilon = 0$ can be written as
\begin{equation} \label{eq:pair-disclination-isotropic-solution}
    \theta_\text{iso}
    =
    q_1 \varphi_1 (x, y)
    + q_2 \varphi_2 (x, y)
    + \frac{\pi}{2}
\end{equation}
where $q_1, q_2$ are the corresponding disclination charges, and we have introduced polar coordinates $(r_i, \varphi_i)$ centered at each defect position $(x_i, y_i)$ (see Fig. \ref{fig:perturbative-disclination-pair-polar-coordinates} for a diagram of the relevant coordinates).
The constant term rotates the director everywhere by $\pi / 2$, a transformation under which Eq. \eqref{eq:2D-frank-energy} is invariant. For $q_1$ and $q_2$ half integers of opposite sign, this solution and the corresponding one without the constant term are so-called ``isomorphs'', characterized by whether the line connecting the two defects is parallel or perpendicular to the far-field director. For example, with $q_1 = +1/2$ and $q_2 = -1/2$, Eq. \eqref{eq:pair-disclination-isotropic-solution} is the perpendicular isomorph.

By expanding $\theta(x, y) = \theta_\text{iso}(x, y) + \epsilon \theta_c(x, y) +
\mathcal{O}(\epsilon^2)$, and substituting into Eq.
\eqref{eq:frank-cartesian-euler-lagrange} we find a Poisson equation for the
first order correction $\theta_c$: 
\begin{eqnarray} 
\nabla^{2} \theta_{c} & = &
        \frac{q_{1}(2-q_{1})}{r_{1}^{2}} \sin (2(1-q_{1})\varphi_{1}-2 q_{2} 
\varphi_{2})  \nonumber \\
        & + & \frac{q_{2}(2-q_{2})}{r_{2}^{2}} \sin (2(1-q_{2})\varphi_{2}-2 
q_{1} \varphi_{1}) \\
        & - & \frac{2 q_{1}q_{2}}{r_{1}r_{2}} \sin \left[ (1-2q_{1})\varphi_{1}+ (1 -2 q_{2}) \varphi_{2} \right]  
\label{eq:perturbative-disclination-pair-poisson}
\end{eqnarray}
We point out that the other isomorph merely changes the right-hand side --and therefore the solution-- by a sign. In what follows, we find an approximate solution to Eq. \eqref{eq:perturbative-disclination-pair-poisson} in various regions which can then be compared against numerical results.

For concreteness, we choose $q_1 = +1/2$ and $q_2 = -1/2$. Near one of the disclinations, $(x_1, y_1)$, one may rewrite $\varphi_2$ and $r_2$ in terms of $\varphi_1$ and $r_1$. In this region, $r_1 / d \ll 1$ so that we Taylor expand the right-hand side to find,
\begin{equation} 
    \nabla^2 \theta_c
    =
    -\frac{3}{4 r_1^2} \sin\varphi_1
    + \frac{3}{8 d r_1} \sin 2\varphi_1
    + \mathcal{O}\left( \frac{r_1}{d} \right)
\end{equation}
A particular solution $\theta^{p, 1}_c$ can be found as given by
\begin{equation}\label{eq:plus-half-perturbative-solution}
    \theta^{p, 1}_c
    =
    \frac34 \sin \varphi_1 
    - \frac{r_1}{8 d} \sin 2\varphi_1
\end{equation}
By comparing it with Eq. \eqref{eq:isolated-defect-perturbative-solution}, we note that the term independent of $r_1$ corresponds to the correction for an isolated disclination in an anisotropically elastic medium, while the term due to pairwise disclination interaction is new and goes linearly in $r_1$ close to $q_1$. A similar calculation for the region close to $q_2$ yields a particular solution,
\begin{equation}\label{eq:minus-half-perturbative-solution}
    \theta^{p, 2}_c
    =
    \frac{5}{36} \sin 3\varphi_2
    +   
    \frac{r_2}{24d} \left( \sin 2\varphi_2 - \sin 4\varphi_2 \right)
\end{equation}
Again we obtain a term independent of $r_{2}$ which is identical to eq. \eqref{eq:isolated-defect-perturbative-solution}, and an interaction term which is linear in $r_2$.

Finally, in the far-field, one may rewrite the equation to first order in polar coordinates whose origin is midway between the two defects $(r, \varphi)$. Expanding the inhomogeneous term in $d / r \ll 1$ yields,
\begin{equation}
    \nabla^2 \theta_c
    =
    -\frac{2 d}{r^3} \sin 3 \varphi
    + \mathcal{O} \left( \left(\frac{d}{r}\right)^2 \right)
\end{equation}
A particular solution to second order is given by,
\begin{equation}
    \theta^{p, f}_c
    =
    \frac{d}{4 r} \sin 3\varphi
\end{equation}
The dependence on $3 \varphi$ and proportional to $d/r$ at long distances is unexpected. Consider the isotropic solution Eq. \eqref{eq:pair-disclination-isotropic-solution}, and express it in terms of the midpoint polar coordinates,
\begin{align*}
    \theta_\text{iso}
    &=
    q_1 \arctan \left( \frac{\sin \varphi}{\cos \varphi + \frac12 \frac{d}{r}} \right)
    + q_2 \arctan \left( \frac{\sin \varphi}{\cos \varphi - \frac12 \frac{d}{r}} \right) \\
    &= - \frac{d\left(q_{1} - q_{2}\right)}{2 r} \sin\left(\varphi \right)
    + q_1 + q_2 + \mathcal{O}\left( \left(\frac{d}{r} \right)^2 \right)
\end{align*}
If $q_1 + q_2 =0$ the constant terms identically vanishes (charges mutually screen), and the dipolar term has the expected dependence in $d/r \sin \varphi$ from a multipolar expansion. However, anisotropic elasticity changes charge screening, and it introduces a new term that, while also decaying as $d/r$ at long distances, it has a different angular dependence.

A general solution which matches the particular solutions in the inner and far field regions would also require the general solution to Laplace's equation. Far from the disclination pair, one would have,
\begin{equation}
    \theta^{s, f}_c = \sum_n \frac{B_n}{r^n} \sin (n \varphi)
\end{equation}
The inner solutions include the components $n = 1, n = 2, n = 3$, and (although much smaller in magnitude as we will argue below) $n = 4$ components. Hence, we would expect those Fourier modes to be present in the far field in order to match at the near-field far-field boundary, giving an approximate far-field solution of:
\begin{equation} \label{eq:full-perturbative-far-field-solution}
    \theta^f_c 
    \approx
    \frac{d}{4 r} \sin 3\varphi
    + \sum_{n = 1}^4 \frac{B_n}{r^n} \sin n \varphi
\end{equation}
We will not pursue this analytic expansion further. Rather we will argue that this dependence is consistent with our numerical solutions for weak elastic anisotropy shown below.

\section{Numerical solutions for a disclination pair}

\subsection{Director representation} \label{numerical-solution-of-perturbative-director}

Equation \eqref{eq:perturbative-disclination-pair-poisson} is a Poisson equation in which the soource term is singular at the location of the two disclinations. We have modified a preexisting deal.II library program to solve it \cite{re:dealii9.4, re:dealii2019design}. The actual linear system is solved with the conjugate gradient method with Trilinos ML algebraic multigrid as a preconditioner \cite{re:sala2004ml}.  As was the case with the $\mathbf{Q}$ tensor, we take as outer boundary condition a zero normal component of the configurational force, where here the configurational force is $\partial f_{n} / \partial \left( \nabla \theta \right)$ with $f_{n}$ the Frank elastic energy density. Because the solution is found perturbatively, the boundary conditions must be specified order by order (see Appendix \ref{disclination-pair-director-appendix}, Eq. \eqref{eq:perturbative-director-numerical-boundary-conditions} for details). We solve on a circular domain radius $R = 5,500$ and defect spacing $d = 60$. These dimensions are arbitrary, since Eq. \eqref{eq:perturbative-disclination-pair-poisson} remains invariant under a change in length scale, but they have been chosen to correspond with the $\mathbf{Q}$-tensor configuration solution shown later which \textit{does} have a length-scale dependence.

We also solve Eq. \eqref{eq:perturbative-disclination-pair-poisson} inside a modified circular domain that excludes the singular points in its right hand side. We cut out two small discs around each disclination, and impose Dirichlet boundary conditions on the circumference of each discs. For simplicity, we prescribe $\theta_c = 0$ on these internal boundaries which corresponds to $\theta = \theta_\text{iso}$ from eq. \eqref{eq:pair-disclination-isotropic-solution}. We choose the cutout radius $r_\text{cutout} = 10$ because, as evidenced in Figs. \ref{fig:minus-half-disclination-director-angle} and \ref{fig:plus-half-disclination-director-angle}, an isolated disclination in the $Q$-tensor formulation becomes uniaxial with approximately constant-$S$ at approximately $r = 10$. The choice of domain is motivated by the comparison carried out below with a full numerical solution in the $\mathbf{Q}$ representation with the same value of the anisotropy parameter $\epsilon$. In the $Q$ tensor formulation, the configuration with two disclinations is not stationary, and hence allowing an unconstrained configuration relax leads to disclination annihilation. This would prevent us from determining the constrained equilibrium configuration corresponding to two immobile disclinations. 

\begin{figure*}[t]
\captionsetup[subfigure]{justification=Centering}

  \begin{subfigure}[c]{0.45\textwidth}
    \includegraphics[width=\textwidth]{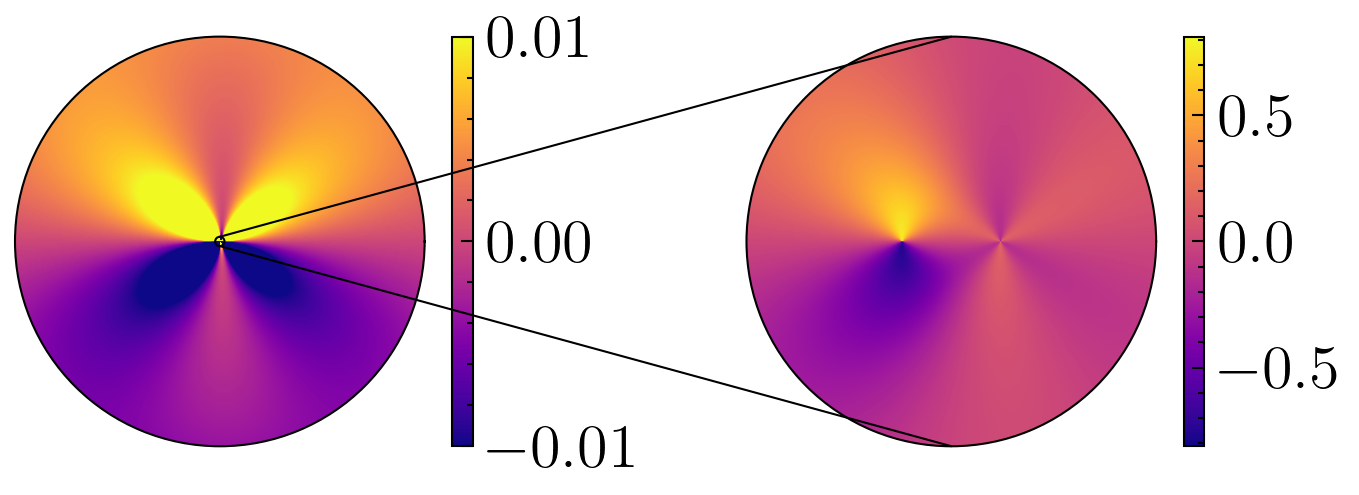}
    \caption{}
    \label{fig:perturbative-disclination-pair-colormap-no-cutout}
  \end{subfigure}
  \begin{subfigure}[c]{0.45\textwidth}
    \includegraphics[width=\textwidth]{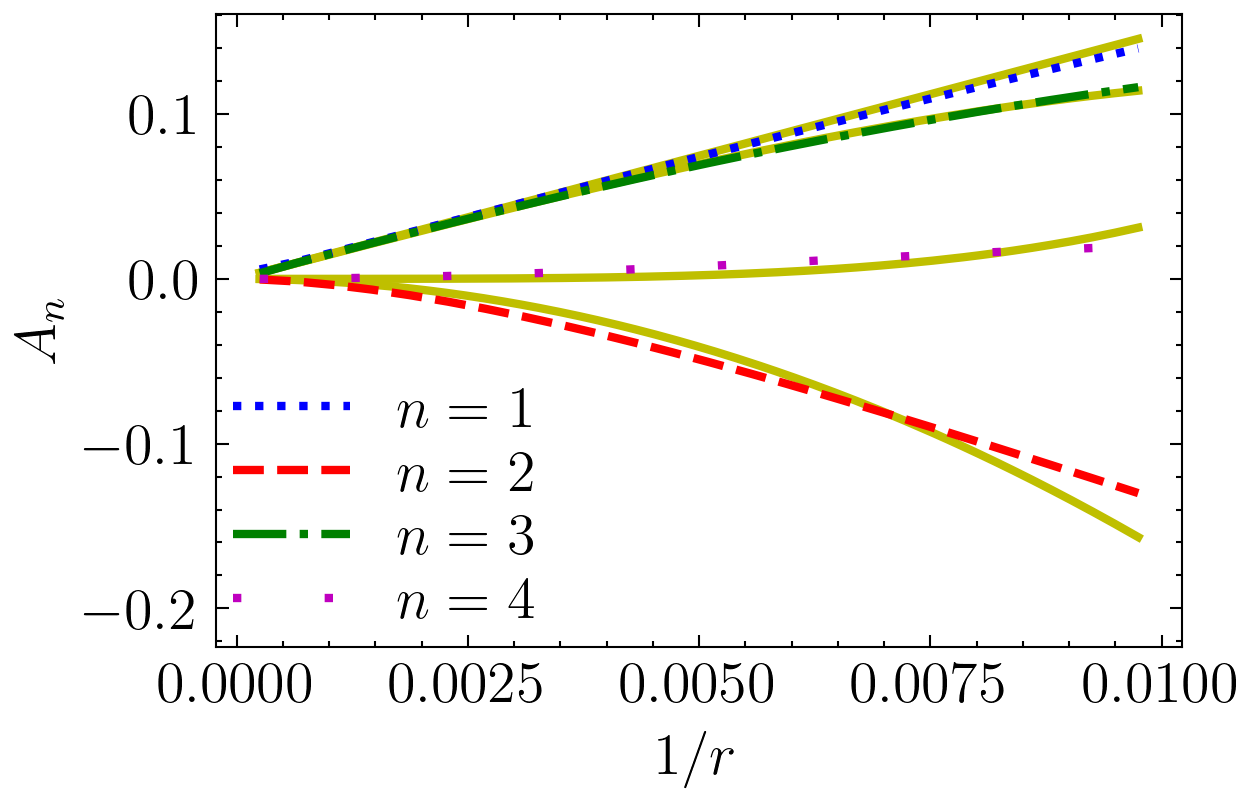}
    \caption{}
    \label{fig:perturbative-disclination-pair-far-field-fourier-no-cutout}
  \end{subfigure}\hfill

  \begin{subfigure}[c]{0.45\textwidth}
    \includegraphics[width=\textwidth]{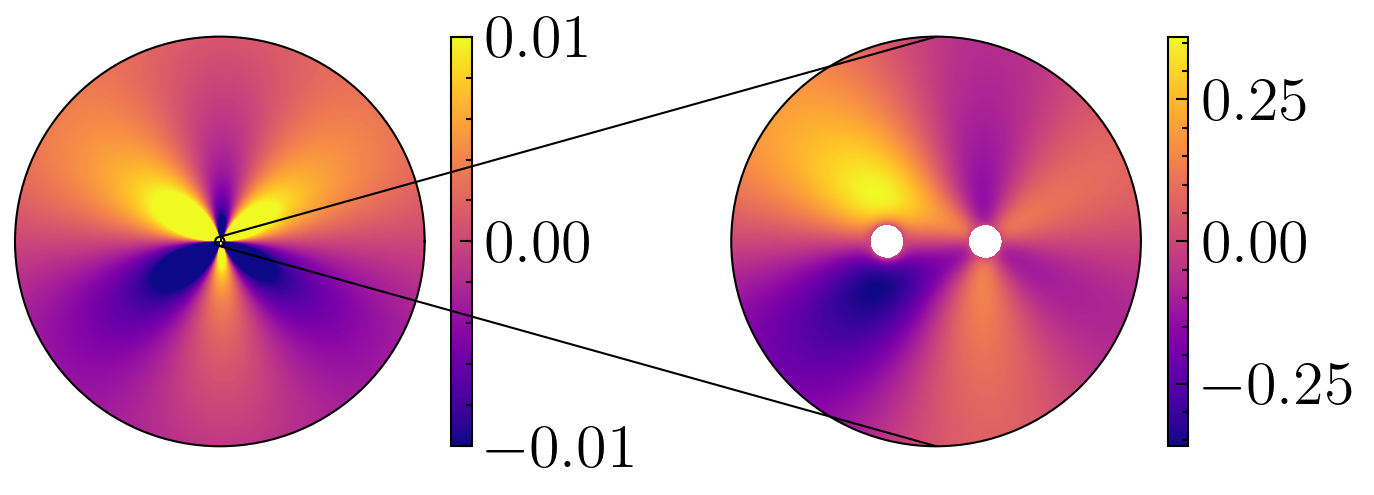}
    \caption{}
    \label{fig:perturbative-disclination-pair-colormap-cutou}
  \end{subfigure}
  \begin{subfigure}[c]{0.45\textwidth}
    \includegraphics[width=\textwidth]{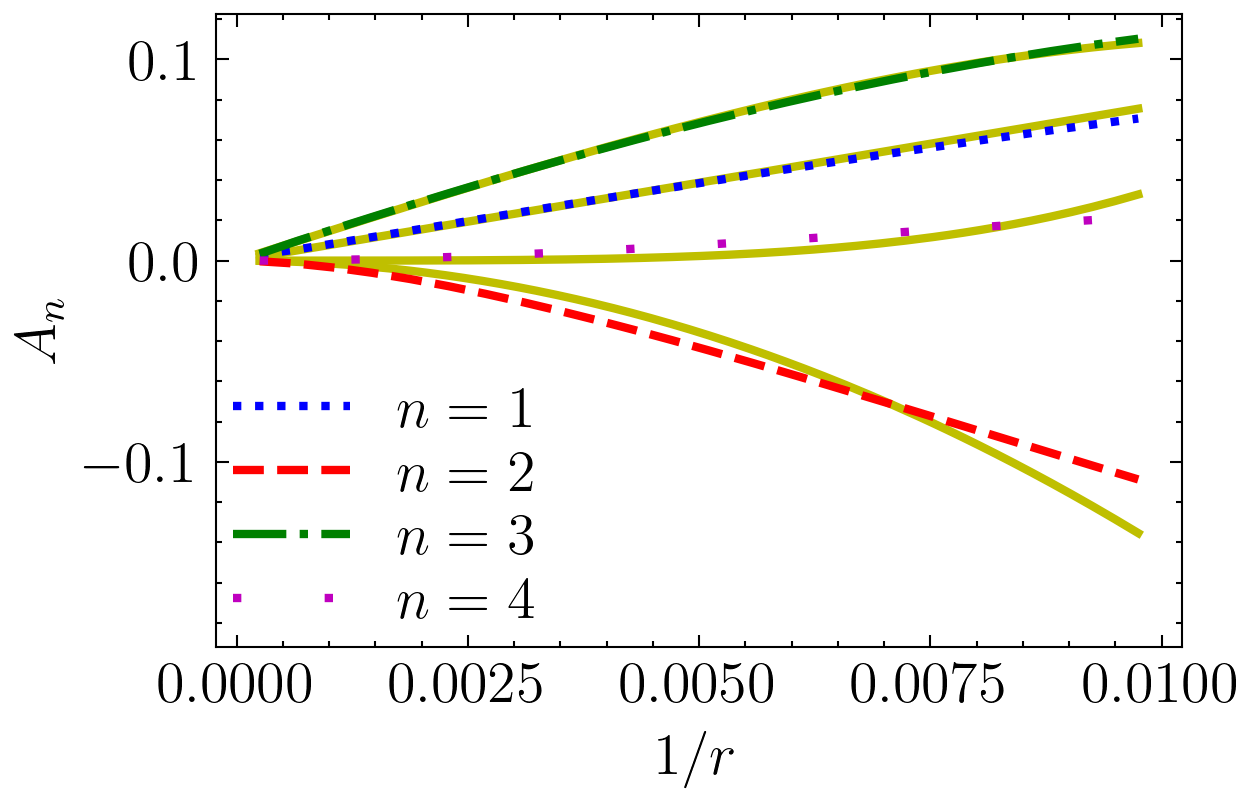}
    \caption{}
    \label{fig:perturbative-disclination-pair-far-field-fourier-cutout}
  \end{subfigure}
  \hfill
    \caption{Perturbation director contribution $\theta_{c}$ corresponding to disclination pair in an elastically anisotropic medium. (a), (c) Colormap of $\theta_c$ in the far field (left) and magnified near field (center). The outer domain radius is $5,500$, while near field magnified region width is $125$. Figure (c) has cut out in the solution domain of radius $10$ around each disclination. $\theta_c = 0$ fixed on their boundaries. (b), (d) Corresponding lowest Fourier modes of $\theta_c$ as a function of $1/r$ in the far field. Curve fits are polynomials with degrees suggested by Eq. \eqref{eq:full-perturbative-far-field-solution}, and are represented by solid lines.}
  \label{fig:perturbative-disclination-pair}
\end{figure*}

Figure \ref{fig:perturbative-disclination-pair-colormap-no-cutout} shows a colormap of $\theta_c$, both in the far field and near field limits. Near the disclination cores one may clearly see the $n = 1$ and $n = 3$ mode contributions from Eqs. \eqref{eq:plus-half-perturbative-solution} and \eqref{eq:minus-half-perturbative-solution} around the +1/2 and -1/2 disclinations respectively. The far field appears to have six fold symmetry, consistent with a contribution from $n = 3$. In order to quantify the contribution from that various Fourier components to $\theta_{c}$, we decompose the far field numerical solution into angular Fourier modes,
\begin{equation}
    \theta_c^f(r, \varphi)
    =
    \sum_n A_n(r) \sin (n\varphi)
\end{equation}
and fit each mode $A_n(r)$ by a polynomial in $1 / r$, with a degree consistent Eq. \eqref{eq:full-perturbative-far-field-solution}. For example, $A_3$ is allowed to have degree 1 and 3 in $1/r$, while $A_2$ is only allowed to have degree 2. Figure \ref{fig:perturbative-disclination-pair-far-field-fourier-no-cutout} shows the angular Fourier coefficients and the corresponding fits. Both the $n = 1$ and $n = 3$ Fourier modes are consistent with the prediction, while the $n = 2$ and $n = 4$ modes deviates somewhat from the expected quadratic and quartic behavior. The linear dependence of the $n = 3$ mode matches the prediction from eq. \eqref{eq:full-perturbative-far-field-solution} in both magnitude and sign.

The effect of adding cutouts to the integration domain around disclination cores is to suppress the near field $n = 1$ and $n = 3$ mode contributions, as can be seen in Fig. \ref{fig:perturbative-disclination-pair-far-field-fourier-cutout}. This reduction translates in the far field into a small reduction in the magnitude of the $n = 3$ mode, and a noticeable reduction in the amplitude of the $n = 1$ mode.

In agreement with the perturbative calculation of Sec. \ref{sec:perturbative_solution}, these numerical results show a different angular dependence of the director angle that arises from disclination interactions in an anisotropic medium. The $n = 3$ Fourier mode decays at the same rate with distance as the $n = 1$ mode arising from the isotropic solution, although it is a factor of $\epsilon / 2$ in magnitude smaller. Depending on the value of the anisotropy parameter, this term could introduce a significant deviation relative to the isotropic interaction terms, and must therefore be considered in, for example, disclination ensemble dynamics in elastically anisotropic media. Note also that the sign of the $n=3$ far field term changes under the transformation to a different disclination pair isomorph. Hence, it is possible that the effective contribution from elastic anisotropy could be smaller in an ensemble of defects containing a distribution of isomorphs.

\subsection{{\bf Q} tensor representation} 
\label{numerical-solution-of-Q-tensor-pair-disclination}

With our choice of elastic terms, Eq. \eqref{eq:LdG-elastic-free-energy}, elastic anisotropy is determined by the coefficients $L_2$ and $L_3$ while the Frank elastic anisotropy is solely determined by $\epsilon$. Given Eq. \eqref{eq:elastic}, we focus on $L_2 = 0$ and find that $L_3 = 0.3065$ for $\epsilon = 0.1$, a regime in which Eq. \eqref{eq:perturbative-disclination-pair-poisson} should hold. We consider a disc of radius $R = 5,500$, defect spacing $d = 60$, and defect cutout radius $r_\text{cutout} = 10$. The Maier-Saupe constant $\kappa = 8.0$, which corresponds to an equilibrium value of $S_0 = 0.6751$.

\begin{figure}[t]
    \includegraphics[width=0.45\textwidth]{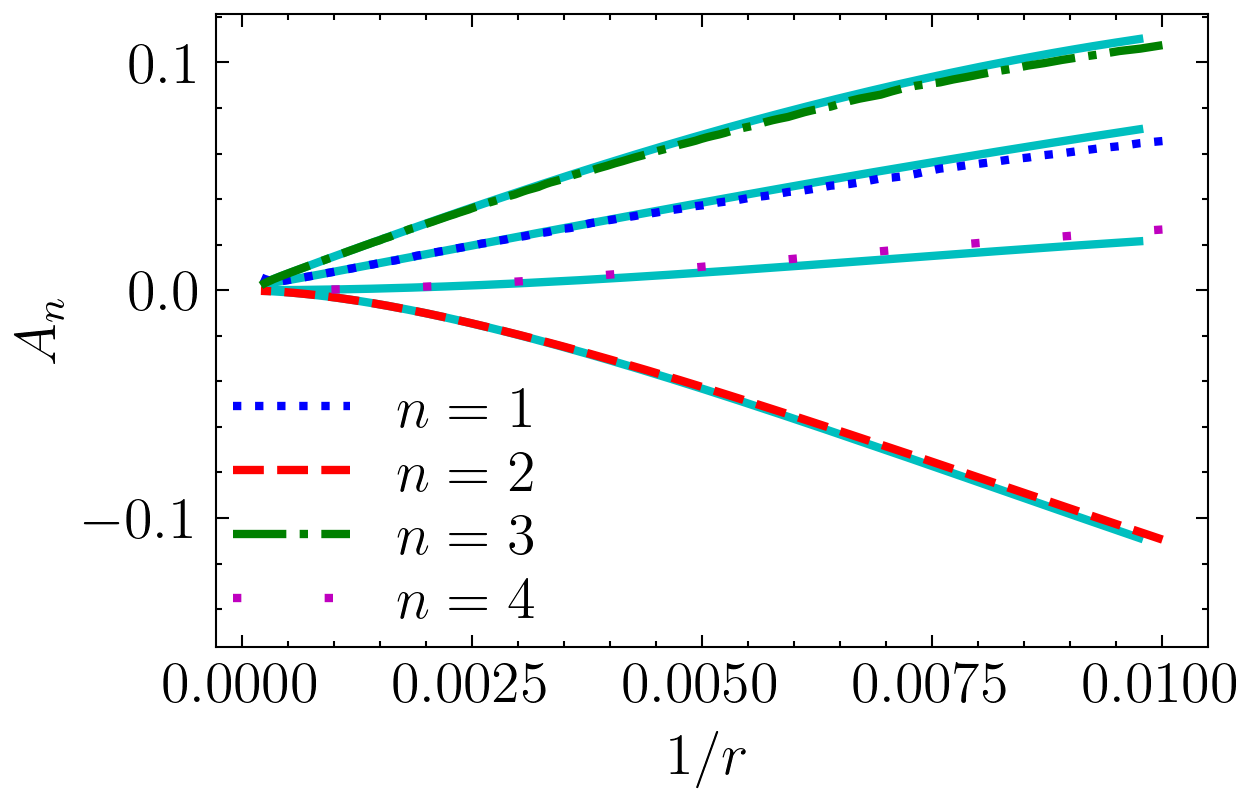}
    \caption{Dotted and dashed lines: Far field angular Fourier components of the eigenvector angle of $\mathbf{Q}$ for the largest eigenvalue (the uniaxial director from $\mathbf{Q}$). For the purposes of the comparison, the isotropic solution \eqref{eq:pair-disclination-isotropic-solution} has been substracted. Solid lines: numerical solution of Eq. \eqref{eq:perturbative-disclination-pair-poisson} (in the director representation.}
    \label{fig:Q-tensor-theta-c-comparison}
\end{figure}

Because of the large size of the computational domain, a direct solution of the minimization problem (Eq. \eqref{eq:Q-tensor-equation-of-motion} with $\partial_{t} \mathbf{Q} = 0$) is difficult. We instead iterate Eq. \eqref{eq:Q-tensor-equation-of-motion} in time until a stationary configuration is reached. As initial condition we choose,
\begin{equation}
    \mathbf{Q}(t=0)
    =
    R(\theta_c) \, \mathbf{Q}_\text{iso} \, R^T(\theta_c)
\end{equation}
%
%
where $R$ is a rotation matrix about the $\mathbf{\hat{z}}$ axis by angle $\theta_c$, which is the numerical solution to Eq. \eqref{eq:perturbative-disclination-pair-poisson} with disclination cutouts fixed at zero. We define $\mathbf{Q}_\text{iso} = S(r_1, r_2) \left( \mathbf{\hat{n}}_\text{iso} \otimes \mathbf{\hat{n}}_\text{iso} - \tfrac13 I \right)$ with $S(r_1, r_2) = S_0 \left( \frac{2}{1 + e^{-r_1}} + \frac{2}{1 + e^{-r_2}} - 3 \right)$ and $\mathbf{\hat{n}} = \begin{bmatrix} \cos \theta_\text{iso} &\sin\theta_\text{iso} &0\end{bmatrix}^T$.
Figure \ref{fig:Q-tensor-theta-c-comparison} shows $\theta_c$ as calculated from the $\mathbf{Q}$ tensor representation compared to $\theta_c$ from Eq. 
(\eqref{eq:perturbative-disclination-pair-poisson}) within the cutout domain.

%
%

%

\section{Conclusions}

We have presented an analysis of the radial and angular dependencies of the orientation order parameter around both an isolated disclination and a disclination dipole in an elastically anisotropic nematic. In the former case, a singular potential theory in the $\mathbf{Q}$ tensor order parameter representation of the nematic shows that the order parameter approaches isotropy near the core: The eigenvalues of the $Q$ tensor become axisymmetric, in agreement with the elastically isotropic case. We provide a scaling law which shows that the zeroth order angular Fourier of the retardance goes to zero linearly with the radial distance $r'$, while the next order Fourier mode decreases quadratically.

For the case of a disclination dipole, we have presented analytical perturbative solutions in the director representation in the limit of weak anisotropy (small elastic constant contrast $\epsilon$). Solutions are given for the nematic orientation angle both near one of the disclinations in the dipole, and in the far field. Particularly noteworthy is the far field dependence in which the $n = 1$ angular Fourier mode of the isotropic limit is supplemented by an $n = 3$ mode as a leading order term due to anisotropy. The predictions agree very well with numerical calculations in both the director field and $Q$-tensor representations of the nematic.

\section*{Acknowledgements} 

This research has been supported by the National Science Foundations under contracts DMR-1838977 and DMR-2223707. We also acknowledge funding from the National Science Foundation REU program, contract No. PHY-2049645, and by the Extreme Science and Engineering Discovery Environment (XSEDE), which is supported by the National Science Foundation under Grant No. ACI-1548562. 

\appendix

\section{Numerical method for an isolated disclination in the director representation} 
\label{isolated-disclination-director-numerics}

The numerical solution of Eq. \eqref{eq:frank-polar-euler-lagrange}, the one dimensional profile of the director $\theta$, as a function of polar angle $\varphi$ is computed by using the Finite Element framework deal.II \cite{re:dealii9.4, re:dealii2019design}. The equation is solved by iteration with a Newton-Rhapson method on the domain $\varphi \in [0, 2\pi]$. The endpoints are fixed at 0 and $2 \pi q$ to maintain azimuth continuity. The equation residual is defined as,
\begin{equation}
    R(\theta)
    =
    \frac{d^2 \theta}{d \varphi}
    - 
    \epsilon \biggl[ \frac{d^{2}\theta}{d \varphi^{2}} \cos 2 (\theta - \varphi) 
        + \left( 2 \frac{d \theta}{d \varphi} - \left( \frac{d \theta}{d \varphi}\right)^{2} \right) \sin 2 (\theta - \varphi)  
    \biggr]
\end{equation}
A Gateaux derivative is introduced,
\begin{equation}
    \begin{split}
        dR(\theta) \delta \theta
        &=
        \left. \frac{d}{d \lambda} R(\theta + \lambda \, \delta \theta) \right|_{\lambda = 0} \\
        &= 
        \frac{d}{d \varphi} \left( p(\varphi) \frac{d \, \delta \theta}{d \varphi} \right)
        + \left( q_1(\varphi) + \frac{d}{d \varphi} q_2(\varphi) \right) \, \delta \theta
    \end{split}
\end{equation}
with 
\begin{equation}
    \begin{split}
        p(\varphi)
        &=
        1 - \epsilon \cos 2 \left( \theta - \varphi \right) \\
        q_1(\varphi)
        &=
        \left[
            \left( \frac{d\theta}{d\varphi} \right)^2
            - 4 \frac{d \theta}{d\varphi}
            + 2
        \right]
        2 \epsilon \cos 2 \left(\theta - \varphi \right) \\
        q_2(\varphi)
        &=
        \frac{d \theta}{d \varphi} 2 \epsilon \sin 2 \left(\theta - \varphi \right)
    \end{split}
\end{equation}
Also define:
\begin{equation}
    q_3(\varphi)
    =
    \left[ 
        \left(\frac{d \theta}{d \varphi} \right)^2 
        - 2
    \right]
    \sin 2 \left(\theta - \varphi \right)
\end{equation}
so that we may write the residual as:
\begin{equation}
    R(\theta)
    =
    \frac{d}{d \varphi} p(\varphi)
    - \left( q_2(\varphi) + q_3(\varphi) \right)
\end{equation}
An iteration in Newton-Rhapson method then reads:
\begin{equation}
    \begin{split}
        dR\left(\theta^{(n)} \right) \, \delta \theta^{(n)}
        =
        - R\left(\theta^{(n)}\right) \\
        \theta^{(n + 1)}
        =
        \theta^{(n)} + \alpha \, \delta \theta^{(n)}
    \end{split}
\end{equation}
with damping parameter $\alpha \leq 1$.
To solve with the finite element method, we take the inner product with a test function $\eta$ and integrate by parts:
$$
  \left< \eta, dR (\theta) \, \delta \theta\right> = 
-\left< \eta, R(\theta) \right> 
$$
$$
        \implies
            \left< \eta, \frac{d}{d\varphi} \left( p \frac{d \, \delta \theta}{d \varphi} \right)
            + \frac{d q_2}{d \varphi} \delta \theta \right>
            + \left< \eta, q_1 \delta \theta \right> 
            =
            -\left< \eta, \frac{dp}{d\varphi} \right>
            -\left< \eta, q_2 + q_3 \right> 
$$
\begin{equation}
        \implies
            -\left< \frac{d \eta}{d \varphi}, p \frac{d \, \delta \theta}{d \varphi} + q_2 \, \delta \theta \right>
            -\left< \frac{d\eta}{d\varphi}, q_2 \frac{d \, \delta \theta}{d \varphi} \right>
            + \left< \eta, q_1 \, \delta \theta \right> \\
            =
            \left< \frac{d \eta}{d \varphi}, p \right>
            - \left< \eta, q_2 + q_3 \right>
\label{eq:isolated-disclination-weak-form}
\end{equation}
The test functions are zero on the boundaries so that the surface integrals vanish.
Approximating $\delta \theta = \sum_j \delta \theta_j \, \eta_j$ with test functions $\eta_j$ given by piecewise polynomial Lagrange elements, and enforcing eq. \eqref{eq:isolated-disclination-weak-form} for each test function $\eta_i$ gives a linear system in $\delta \theta_j$.
We iterate until the $L_2$ norm of the residual is less than some desired threshold.
For the simulations run in this paper, the domain is broken into $2^{10}$ evenly-spaced segments, we use first degree Lagrange elements, and the residual $L_2$ norm tolerance is set to $10^{-10}$. We use the UMFPACK direct sparse matrix solver since, in one dimension at this size, performance is not an issue.

\section{Numerical method in the $Q$-tensor representation} 
\label{Q-tensor-numerics-appendix}

In order to solve Eq. \eqref{eq:Q-tensor-equation-of-motion} numerically we also use the deal.II finite element framework \cite{re:dealii9.4, re:dealii2019design}. This library has the benefit of implementing adaptive mesh refinement, as well as being massively parallelizeable via MPI, allowing for very large scale computations. To solve all linear systems in our implementation, we use the Trilinos linear algebra library via deal.II \cite{re:trilinos-website}. The code developed is available in the GitHub repository \cite{re:lucas_code}. To integrate eq. \eqref{eq:overdamped-equation-of-motion}, consider that the variation of the free energy is given explicitly by:
\begin{equation}
\begin{split}
    \delta F (Q, \nabla Q)
    &=
    \left.\frac{d}{d \tau} F(Q + \tau \, \delta Q, \nabla Q + \tau \nabla \delta Q)\right|_{\tau = 0} \\
    &= 
    \int_\Omega \left[ \frac{\partial f}{\partial Q} \delta Q + \frac{\partial f}{\partial  \left(\nabla Q \right)} \nabla \delta Q \right] dV \\
    &=
    \int_\Omega \left[ \frac{\partial f}{\partial Q} - \nabla \cdot \frac{\partial f}{\partial \left(\nabla Q \right)} \right] \delta Q dV
    + \int_{\partial \Omega} \left[ \boldsymbol\nu \cdot \frac{\partial f}{\partial  \left( \nabla Q \right)} \right] \delta Q dS
\end{split}
\end{equation}
where $f$ is the free energy density.
Here we take $\boldsymbol\nu \cdot \partial f / \partial \left(\nabla Q \right) = 0$ as a boundary condition which corresponds to zero normal configuration force at the boundary.
Additionally, to ensure that $\partial_t F \leq 0$ always, we must take:
\begin{equation}
    \frac{\partial Q}{\partial t}
    =
    -\frac{\partial f}{\partial Q} + \nabla \cdot \frac{\partial f}{\partial \left(\nabla Q \right)}
\end{equation}
One may understand this as taking the time evolution in the direction of the variation $\delta Q$ where the variation is chosen to make $\delta F$ negative definite.
To simplify the exposition, take $T^Q = - \partial f / \partial Q$ and $T^{\nabla Q} = \partial f / \partial (\nabla Q)$.
Finally, $T = T^Q + \nabla \cdot T^{\nabla Q}$.
These are given explicitly by:
\begin{align}
    T^Q_{ij}
    &=
    \kappa Q_{ij}
    - \Lambda_{ij}
    - L_3 \left( 
        \left( \partial_i Q_{kl} \right) \left( \partial_j Q_{kl} \right)^T
        - \tfrac13 \left| \partial_k Q_{lm} \right|^2 \delta_{ij}
    \right) \\
    T^{\nabla Q}_{kij}
    &=
    \partial_k Q_{ij}
    + L_2 \left(
        \partial_i Q_{jk}
        + \partial_j Q_{ik}
        - \tfrac23 \partial_l Q_{lk} \delta_{ij}
    \right)
    + 2 L_3 Q_{kl} \partial_l Q_{ij}
\end{align}
We note that the divergence is contracted over the $k$ index.

To discretize eq. \eqref{eq:Q-tensor-equation-of-motion} in time, we use a Crank-Nicolson method:
\begin{equation}
    \frac{Q - Q_0}{\delta t}
    =
    \tfrac12 \left( T + T_0 \right)
\end{equation}
where  $Q_0$ and $Q$ are the $Q$-configurations at the previous and current timesteps respectively, $\delta t$ is the timestep, and $T$ and $T_0$ are evaluated at $Q$ and $Q_0$ respetively.
Because $T$ is nonlinear, we define a residual:
\begin{equation}
    R
    =
    Q - Q_0
    - \tfrac12 \delta t \left( T_0 + T \right)
\end{equation}
To solve for the configuration when $R = 0$, we use a Newton-Rhapson method.
The Gateaux derivative then reads:
\begin{equation}
    dR \, \delta Q =
    \left.
    \frac{d}{d \tau} R(Q + \tau \, \delta Q, \nabla Q + \tau \nabla \delta Q) \right|_{\tau = 0} = \delta Q - \tfrac12 dT \, \delta Q
\end{equation}
Explicitly, this yields:
\begin{equation}
    \left(dT^Q \, \delta Q\right)_{ij} = \kappa \delta Q_{ij}
        - d\Lambda_{ij}
        - L_3 \biggl(
            \left( \partial_i \delta Q_{kl} \right) \left( \partial_j Q_{kl} \right) 
            +  \left( \partial_i Q_{kl} \right) \left( \partial_j \delta Q_{kl} \right)
            - \tfrac23 \left( \partial_k Q_{lm} \right) \left( \partial_k \delta Q_{lm} \right) \delta_{ij}
        \biggr)
\end{equation}
\begin{equation}
    \left(dT^{\nabla Q} \, \delta Q\right)_{kij} =
        \partial_k \delta Q_{ij}
        + 2 L_3 \left( \delta Q_{kl} \partial_l Q_{ij} + Q_{kl} \partial_l \delta Q_{ij} \right)
\end{equation}
where $d\Lambda_{ij}$ is given by:
\begin{equation}
\begin{split}
    d\Lambda_{ij}
    &=
    \left.\frac{d}{d \tau} \Lambda_{ij}(Q + \tau \, \delta Q)\right|_{\tau = 0} \\
    &=
    \frac{d}{d \tau} \left[ 
        \Lambda(Q) 
        + \tau \frac{\partial \Lambda_{ij}}{\partial Q_k} \delta Q_k 
        + \mathcal{O}(\tau^2)
    \right]_{\tau = 0} \\
    &=
    \frac{\partial \Lambda_{ij}}{\partial Q_k} \delta Q_k
\end{split}
\end{equation}
The Taylor series expansion of $\Lambda$ about $Q$ involves the directional derivative in the direction of $\delta Q$. 
Since $Q$ and $\delta Q$ are restricted to the submanifold of traceless, symmetric tensors, this directional derivative can be accomplished by differentiating $\Lambda$ with respect to the degrees of freedom of $Q$ and dotting into the degrees of freedom of $\delta Q$. 
This set of degrees of freedom is arbitrary, but we note that the space of traceless, symmetric tensors is five-dimensional.
Newton's method then reads:
\begin{equation}
\begin{split}
    &dR \, \delta Q = -R \\
    &Q \to Q + \alpha \, \delta Q
\end{split}
\end{equation}
where we indicate that the next iteration is updated by adding $\alpha \, \delta Q$ with $0 < \alpha \leq 1$ some stabilization constant.

To discretize in space, we find the weak form of this equation by taking the inner product with some symmetric, traceless tensorial test function $\Phi$:
\begin{equation} \label{eq:Q-tensor-weak-form}
    \left< \Phi, dR \, \delta Q \right>
    =
    -\left< \Phi, R \right>
\end{equation}
Approximating $\delta Q$ in our space of test functions gives:
\begin{equation}
    \delta Q = \sum_j \delta Q_j \Phi_j
\end{equation}
where $\delta Q_j$ are a set of scalars, and $\Phi_j$ are a finite element basis.
Asserting that eq. \eqref{eq:Q-tensor-weak-form} be true for a finite number of test functions $\Phi_i$ yields a finite linear system in $\delta Q_j$:
\begin{equation}
    \sum_j \left[ \left< \Phi_i, dT^Q \Phi_j \right> - \left< \nabla \Phi_i, dT^{\nabla Q} \Phi_j \right> \right] \delta Q_j
    =
    \left< \Phi_i, T^Q \right> - \left< \nabla \Phi_i, T^{\nabla Q} \right>
\end{equation}
Note that we have integrated by parts and taken the boundary terms to zero, due to the zero configurational force condition.

In our actual simulations, we take the finite element basis functions $\Phi$ to be piecewise scalar Lagrange polynomials $\phi(\mathbf{x})$ multiplied by constant tensor basis elements $X$:
\begin{equation}
    \begin{split}
    X_1
    =
    \begin{pmatrix}
        1 & 0 & 0 \\
        0 & 0 & 0 \\
        0 & 0 & -1
    \end{pmatrix}
    X_2
    =
    \begin{pmatrix}
        0 & 1 & 0 \\
        1 & 0 & 0 \\
        0 & 0 & 0
    \end{pmatrix}
    X_3
    =
    \begin{pmatrix}
        0 & 0 & 1 \\
        0 & 0 & 0 \\
        1 & 0 & 0
    \end{pmatrix} \\
    X_4
    =
    \begin{pmatrix}
        0 & 0 & 0 \\
        0 & 1 & 0 \\
        0 & 0 & -1
    \end{pmatrix}
    X_5
    =
    \begin{pmatrix}
        0 & 0 & 0 \\
        0 & 0 & 1 \\
        0 & 1 & 0
    \end{pmatrix}
    \end{split}
\end{equation}

In section \ref{numerical-solution-of-Q-tensor-pair-disclination} we use this method with $\delta t = 0.1$, and iterate for $50,000$ time steps.
The tolerance for the residual is an $L_2$ norm of the finite element vector of $1e-10$.
In section \ref{sec:isolated-disclination-core} we instead solve for $\partial Q / \partial t = 0$ to find the equilibrium state.
For this, the zeros of $T$ are found using a Newton-Rhapson method, and the $L_2$ norm tolerance of the residual is $1e-10$.

\section{Numerical method for a disclination pair in the director representation} 
\label{disclination-pair-director-appendix}

Equation \eqref{eq:perturbative-disclination-pair-poisson} is a straightforward Poisson equation, so taking the right-hand side to be $g(x, y)$ we may write the weak form as:
\begin{equation}
    \left< \nabla \phi, \nabla \theta_c \right>
    - \left< \phi, \mathbf{n} \cdot \nabla \theta_c \right>_{\partial \Omega}
    =
    -\left< \phi, g \right>
\end{equation}
where here $\phi$ is a test function, $\left< , \right>$ is the $L^2$ inner product over the domain, and $\left<, \right>_{\partial \Omega}$ is the $L^2$ inner product over the boundary.

Because we cannot solve numerically on an infinite domain, we seek a finite domain and boundary conditions which correspond most closely with our infinite-domain analytic solution.
For both the $Q$-tensor and director model, we enfore zero normal configurational stress:
\begin{equation}
    \mathbf{N} \cdot \frac{\partial f}{\partial (\nabla \theta)} = 0
\end{equation}
where $f$ is the Frank free energy density.
Explicitly, the configurational stress in an anisotropic medium is:
\begin{equation}
    \frac{\partial f}{\partial (\nabla \theta)}
    =
    \nabla \theta
    +
    \epsilon
    C (\theta)
\end{equation}
where we have defined:
\begin{equation}
    C(\theta)
    =
    \begin{bmatrix}
    \sin2\theta \left(\partial_y \theta\right)
    + \cos2\theta \left(\partial_x \theta\right)\\
    \sin2\theta\left(\partial_x \theta\right)
    - \cos2\theta \left(\partial_y\theta\right)
    \end{bmatrix}
\end{equation}
To first order, the zero-configurational stress condition reads:
\begin{equation}
    \nabla \theta_\text{iso}
    + \epsilon \nabla \theta_c
    + \epsilon C(\theta_\text{iso})
    =
    0
\end{equation}
Order by order, we note:
\begin{equation}
    \left.\frac{\partial \theta_\text{iso}}{\partial r} \right|_{r = R}
    =
    \frac{q_1 2 d \sin(\varphi)}{d^2 + 4dR \cos(\varphi) + 4R^2}
    -
    \frac{q_2 2 d \sin(\varphi)}{d^2 - 4dR \cos(\varphi) + 4R^2}
\end{equation}
where $R$ is the radius of the circular domain.
This goes as $d / R^2$, and so goes to zero in the limit that $d / R \ll 1$.
The first order anisotropic correction boundary term then goes as:
\begin{equation} \label{eq:perturbative-director-numerical-boundary-conditions}
    \mathbf{N} \cdot \nabla \theta_c
    =
    - \mathbf{N} \cdot C(\theta_\text{iso})
\end{equation}
Given these two conditions, the zero configurational stress is met up to first order.

For the finite element simulation, we use first order Lagrange elements as test and shape functions, and solve iteratively with the Conjugate gradient method with convergence tolerance $10^{-12}$. 
As a preconditioner, we use the Trilinos ML Algebraic Multigrid method.

\bibliography{main}
\bibliographystyle{unsrt}

\end{document}